# Emerging from The Cloud: A Bibliometric Analysis of Cloud Forensics Studies

**James Baldwin, Omar M. K. Alhawi, Simone Shaughnessy, Alex Akinbi, Ali Dehghantanha**

**Abstract** Emergence of cloud computing technologies have changed the way we store, retrieve, and archive our data. With the promise of unlimited, reliable and always-available storage, a lot of private and confidential data are now stored on different cloud platforms. Being such a gold mine of data, cloud platforms are among the most valuable targets for attackers. Therefore, many forensics investigators have tried to develop tools, tactics and procedures to collect, preserve, analyse and report evidences of attackers' activities on different cloud platforms. Despite the number of published articles there isn't a bibliometric study that presents cloud forensics research trends. This paper aims to address this problem by providing a comprehensive assessment of cloud forensics research trends between 2009 – 2016. Moreover, we provide a classification of cloud forensics process to detect the most profound research areas and highlight remaining challenges.

## 1 Introduction

Cloud Computing is an emerging technology that has seen a rapid adoption by enterprises and individual consumers. Gartner forecasted that the cloud computing market will hit US$250 billion by 2017 as cloud adoption increases in organizations [1]. Cisco have also forecasted that annual global cloud IP traffic will reach 14.1 ZB (14.1 billion TB ) by the end of 2020, up from 3.9 ZB in 2015 [2].

School of Computing, Science and Engineering, University of Salford, Manchester, U.K.

Email: J.Baldwin1@edu.salford.ac.uk, O.Alhawi@edu.salford.ac.uk, S.Shaughnessy@edu.salford.ac.uk, O.A.Akinbi@salford.ac.uk, A.Dehghantanha@salford.ac.uk

The National Institute of Standards and Technology (NIST) considers three cloud service models [3]: Software as a Service (SaaS), Platform as a Service (PaaS) and Infrastructure as a Service (IaaS). In the SaaS model an Application Service Provider (ASP) provides various applications over the Internet which eliminates the need for software and IT infrastructure (servers/databases etc) maintenance for the ASP customers [4]. The applications are accessed using a client browser interface. Google Apps, Yahoo Mail and CRM applications are all instances of SaaS. In the PaaS model, cloud infrastructure is owned and maintained by the provider; the customer is then able to deploy and configure applications into a provider managed framework and infrastructure [5]. Examples of PaaS are Google App Engine, Apprenda and Heroku. In the IaaS model resources are provided to the customer as virtualised resources e.g. Virtual Machines (VMs). Whereas the customer has full control over the operating system, the provider maintains control over the physical hardware. This allows for services to be scaled and billed in line with customer resource requirements [6]. Amazon Web Services (AWS), Microsoft Azure and Google Compute Engine (GCE) are examples of IaaS models.

Furthermore, NIST suggests four cloud deployment models, namely private cloud, community cloud, public cloud, and hybrid cloud. With public clouds, services are available through a public cloud service provider (Microsoft, Amazon, etc) who host the cloud infrastructure, and customers don't have any control over the located infrastructure. Private clouds are dedicated to organizations (as opposed to the public) and host specific business relevant applications. Community clouds are shared between organisations with similar requirements and business objectives; they are maintained by all participating members of the community. The final model, hybrid cloud, consist of 2 or more of the public, private and community models [7].

According to the 2016 State of the Cloud Survey in which 1060 technical professionals representing a broad cross section of organizations were questioned [8], there has been an increase (from 2015) in the number of organisations utilizing the services of cloud providers. This change is illustrated in Figure 1.

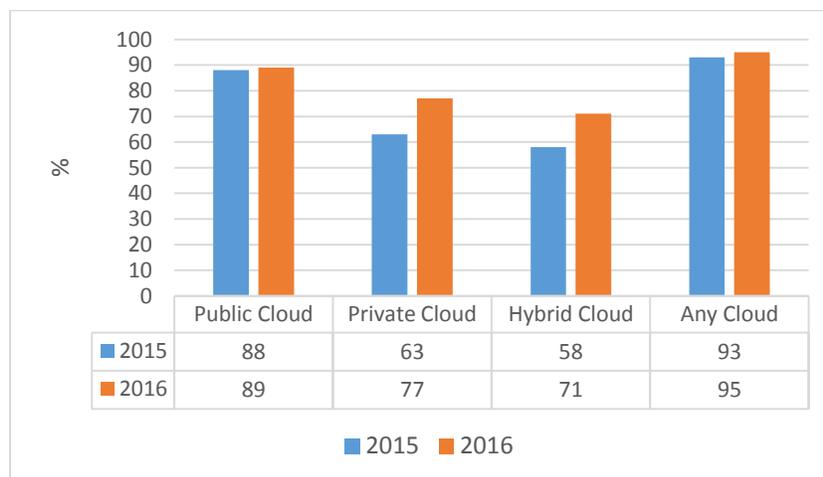

**Figure 1.** Survey Respondents Adopting Cloud - 2016 vs. 2015 [8]

Actual statistical figures of how many crimes have been committed in the cloud are unclear as Cloud Service Providers (CSPs) often ask clients not to disclose any information to the public in relation to cyber incidents [9]. As more organisations move away from traditional 'in house' computing and adopt cloud technology to provide the infrastructure to run their businesses, there are more opportunities and vulnerabilities for attackers in such a rapidly changing environment. According to DarkReading [10] the number of cybercrime incidents reported in the UK has surpassed traditional crime within this current year. With the crime rate increasing, the need for forensic investigations within the cloud has also increased.

The term "cloud forensics" (a cross-discipline of cloud computing and digital forensics [11]) was first introduced in 2011 [11] to recognize the rapidly emerging need for digital investigation in cloud computing environments. According to the 2016 State of the Cloud Survey [8] there is a lack of forensic tools that are tailored for cloud systems. Approximately 58% of respondents agreed that digital forensic process automation is needed to tackle future challenges including cloud forensics [12]. Current forensic tools appear unsuited to process cloud data as the physical inaccessibility of the evidence and lack of control over the system make evidence acquisition a challenging task [13]. Table 1 illustrates a summary of mostly used tools to conduct cloud forensics [14]

**Table 1**
A summary of mostly used tools to conduct cloud forensics [14]

| Utilized tools | General/cloud-based tools | Functionality |
| --- | --- | --- |
| FTK remote agent [15] | General | Remote drive and memory image acquisition; remote mounting |
| Encase remote agent [16] | General | Remote drive image acquisition |
| Snort [17] | General | Network traffic monitoring and packet logging |
| FROST [18] | Cloud-based | Digital forensics tools for the OpenStack cloud platform |
| OWADE [19] | Cloud-based | Reconstruction of browsing history and online credentials |
| CloudTrail [20] | Cloud-based | Logging in the AWS cloud |
| Wireshark [21] | General | Network traffic capture and analysis |
| Sleuthkit [22] | General | Forensic image analysis and data recovery |
| FTK imager [15] | General | Acquisition of memory and disk images |
| X-Ways [23] | General | Acquisition of live systems (Windows and Linux) |
| Encase e-discovery suite [24] | General | Drive image acquisition and offline examination |

Variety of investigation frameworks have been suggested for cloud forensics [25],[26]. Moreover, researchers tried to identify residual evidences of users' interactions with different cloud platforms such as DropBox [27], MEGA [28], GoogleDrive [29], SugarSync [30], pCloud [31], [32], CloudMe [33], SpiderOak [32] and hubiC [34] on Windows, Linux and mobile devices. There were several attempts to extract server-side evidences of different cloud platforms such as Syncany [35], BitTorrent Sync [36], SymForm [37] as well. While there was a lot of focus on technological and procedural development in cloud forensics [38], to the best of authors' knowledge, a bibliometric analysis of this emerging technology does not exist. As such, this paper aims to provide a comprehensive bibliometric analysis of cloud forensics studies and to demonstrate research trends by highlighting the substantial research contributions. We discuss publication statistics, citation distributions and statistics, regional and institutional productivity, research areas, impact journals, and keywords frequency. By identifying research gaps and challenges in the forensic process this paper will open the way for future research within cloud forensics.

This paper is organised as follows: Section 2 describes the research methodology;

Section 3 presents the results and discusses cloud forensics studies; Section 4 introduces the challenges and future trends; Section 5 is the conclusion to the study.

## 2 Methodology

Bibliometrics is a method which allows us to verify the relevance, appropriateness and research impacts of a research area/subject based on citation metrics [39]. According to Eugene Garfield [40] the citation index has a quantitative value which helps to define the significance of an article. This in turn helps to measure the 'influence' or 'impact factor which is based on 2 elements: the number of citations in the current year to items published in the previous 2 years, and the number of substantive articles and reviews published in the same 2 years [41].

We chose to use Web of Science (WoS) as our primary database researching about published articles on 'cloud forensics' as a reliable single source for publications. There are other databases available to search such as Google Scholar and Scopus however WoS provides a more comprehensive and accurate image of the scholarly impact of author [42].

As illustrated in Figure 2 the first step in the data collection process was to use key word searches against the WoS database. We used various general search terms including 'cloud forensic*', forensic* (where * denotes a wildcard character in the search term) and 'cloud investigation' in addition to specific search terms to include platforms, service models and deployment methods. The initial 18,275 results were then refined by excluding other databases such as KCI-Korean Journal Database, Medline and Russian Science Citation. Finally, we refined the results by removing unrelated publications providing a final total of 260 publications directly related to the cloud forensics research area.

Analysis of our results were performed using a combination of the WoS results analysis tools and spreadsheet processing to obtain further detail such as clearly defined geographical region and keyword frequency statistics; they are presented and discussed in section 3.

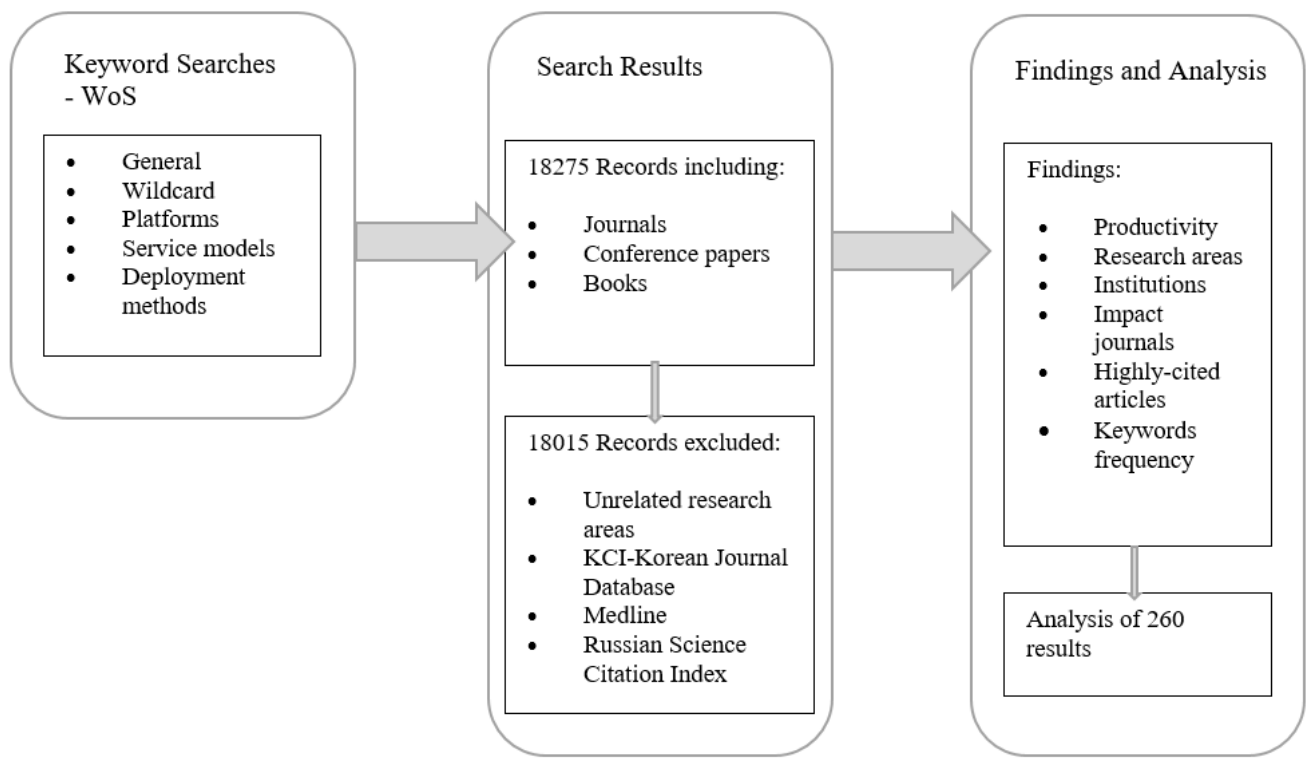

**Figure 2.** The data collection process

## 3  Results and Discussion

Table 2 categorizes related publications based on their source to journals and books (which contain conference proceedings as well). The book category has the highest proportion of publications at 73.08%. This is more than double the journal category which totals 26.92%.

**Table 2**
Publication categories

| Category | No. Publications | % Publications |
|---|---|---|
| Book | 190 | 73.08 |
| Journal | 70 | 26.92 |

The publication frequency of both categories is illustrated in Figure 3.
Between 2011 - 2015 there has been a significant increase in the number of book chapters. In 2011 the introduction of the term "cloud forensics" [11] and the release of both the UK [43] and US [44] cloud computing strategies may have contributed to an increased global focus on cloud computing. This year (2011) also represented the beginning of funding into cloud forensics research with a single (1) publication funded by the National Science Foundation Cyber Trust. From 2011 – 2016 the number of funded articles then increased to 8 with an overall 28 publications funded over this period.

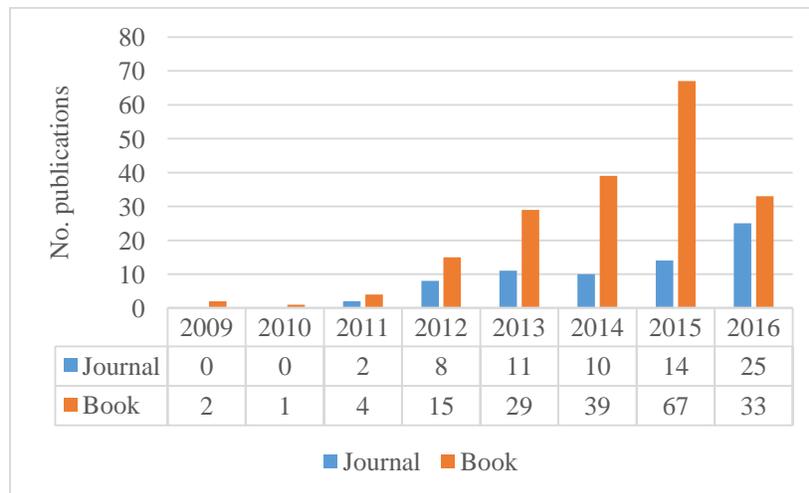

**Figure 3.** Number of publications

WoS provides a citation map feature that allows the researcher to have a holistic view of related research which reflects how researchers embed their work within related and earlier publications [45][46]. Figure 4 shows the annual citation distribution over the period 2009 - 2016. This 7-year period represents the start of research into cloud forensics and the average number of citations over the period is 87.75. In 2011 there was only 1 citation while there were 6 citations in 2012 with significant growth in 2013 (650%), 2014 (305%) and 2015 (172%).

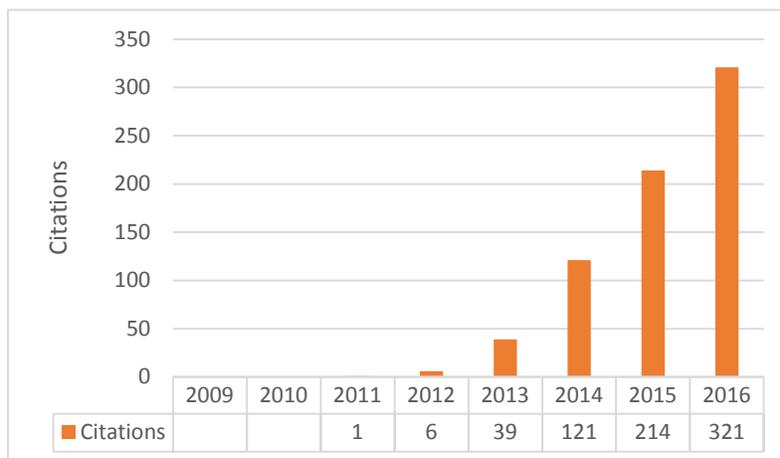

**Figure 4.** Citation distributions

Figure 5 represents the total number of available cloud forensic publications and the number of citations for each year in the period of 2009 to 2016. The earlier an article is published the more citations it received [47]. The top 3 cited publications in this study were published in 2012 which corresponds to the first significant annual citation increase as discussed earlier.

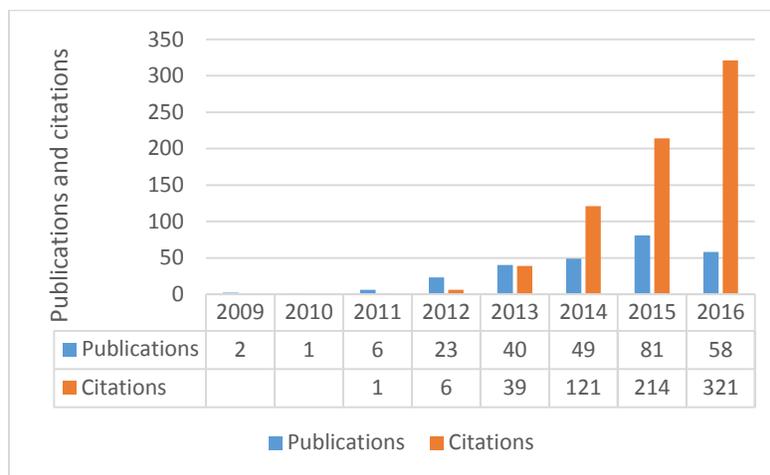

**Figure 5.** Publications and citations

## 3.1 Productivity

This section will discuss productivity based on the publication output from the 6 geographical regions identified in this study. It can show whether there are any significant geographic trends that correlate to institutional and author research contribution.

Table 3 lists the productivity ordered by regional and country contribution. It shows that Asia is the major contributor to cloud forensics publications with a total of 36.92% of all publications. It is closely followed by Europe with 28.85%. The North American, African,

Australian and Middle Eastern regions total 34.23% combined. Asia and Europe are therefore the most productive regions both contributing to almost two-thirds of all publications.

Within Asian India and China are clear research-leaders with a combined 26.92% of overall global publications. 72.92% of all publications produced by 9 countries in Asia. The second highest contributor, Europe, has 2 clear research leaders with England and Ireland contributing 14.23% of all global publications and 49.33% of the publications from the 17 countries in that region. North America is a single leader in the USA with 16.15% overall and 91.30% regional contribution out of just the 2 representative countries. The USA, India and China are the overall top 3 countries with a combined total of 43.08% of all publications. Table 4 shows the major contributing countries within each region.

**Table 3**
Productivity by region

| List of regions | No. of articles | (%) | List of regions | No. of articles | (%) |
| --- | --- | --- | --- | --- | --- |
| **Asia** | **96** | **36.92** | Scotland | 2 | 0.77 |
| India | 40 | 15.38 | Croatia | 1 | 0.38 |
| Peoples R China | 30 | 11.54 | Poland | 1 | 0.38 |
| Taiwan | 7 | 2.69 | Serbia | 1 | 0.38 |
| Japan | 5 | 1.92 | Slovenia | 1 | 0.38 |
| South Korea | 4 | 1.54 | Spain | 1 | 0.38 |
| Bangladesh | 3 | 1.15 | Switzerland | 1 | 0.38 |
| Malaysia | 3 | 1.15 | Wales | 1 | 0.38 |
| Pakistan | 3 | 1.15 | **North America** | **46** | **17.69** |
| Sri Lanka | 1 | 0.38 | USA | 42 | 16.15 |
| **Europe** | **75** | **28.85** | Canada | 4 | 1.54 |
| England | 21 | 8.08 | **Africa** | **22** | **8.46** |
| Ireland | 16 | 6.15 | South Africa | 19 | 7.31 |
| Italy | 7 | 2.69 | Ghana | 1 | 0.38 |
| Greece | 6 | 2.31 | Morocco | 1 | 0.38 |
| Germany | 5 | 1.92 | Tunisia | 1 | 0.38 |
| Romania | 5 | 1.92 | **Australia** | **17** | **6.54** |
| France | 2 | 0.77 | Australia | 17 | 6.54 |
| Netherlands | 2 | 0.77 | **Middle East** | **4** | **1.54** |
| Norway | 2 | 0.77 | U Arab Emirates | 4 | 1.54 |

**Table 4**
Productivity by leading regional countries

| Country | Region | No. countries in region | Publications (%) | Contribution to region % |
| --- | --- | --- | --- | --- |
| USA | North America | 2 | 16.15 | 91.30 |
| India | Asia | 9 | 15.38 | 41.67 |
| Peoples R China | Asia | 9 | 11.54 | 31.25 |
| England | Europe | 17 | 8.08 | 28.00 |
| Australia | Australia | 1 | 6.54 | 100.00 |
| South Africa | Africa | 4 | 7.31 | 86.36 |
| Ireland | Europe | 17 | 6.15 | 21.33 |
| U Arab Emirates | Middle East | 1 | 1.54 | 100.00 |

### 3.2 Research Areas

The WoS database contains 150 different scientific research areas that can be categorised depending on the focus and reach of the research across multiple sectors. As seen in Table 5 they can be a combination of single and multi-disciplined research. Table 5 illustrates that 50% of all publications are attributed to the single-disciplined Computer Science research area. The second and third highest research areas are the multi-disciplined Computer Science & Engineering (21.54%), and Computer Science and Telecommunications (8.46%). Within this study the most influential publications in the top 3 research areas are "Acquiring Forensic Evidence from Infrastructure-As-A-Service Cloud Computing: Exploring and Evaluating Tools, Trust, And Techniques", "Cloud Computing-Based Forensic Analysis for Collaborative Network Security Management System" and "A Cloud Computing Platform for Large-Scale Forensic Computing".

**Table 6**
Research Areas in isolation

| Research Areas | No. Publications | % |
|---|---|---|
| Computer Science | 230 | 88.46 |
| Engineering | 91 | 35.00 |
| Telecommunications | 48 | 18.46 |
| Automation Control Systems | 5 | 1.92 |
| Government Law | 5 | 1.92 |
| Information Science Library Science | 3 | 1.15 |
| Criminology Penology | 2 | 0.77 |
| International Relations | 2 | 0.77 |
| Materials Science | 2 | 0.77 |
| Medical Informatics | 2 | 0.77 |
| Operations Research Management Science | 2 | 0.77 |
| Education Educational Research | 1 | 0.39 |
| Legal Medicine | 1 | 0.39 |
| Physics | 1 | 0.39 |
| Science Technology Other Topics | 1 | 0.39 |

**Table 5**
Research Areas

| Research Areas | Publications | % |
| --- | --- | --- |
| Computer Science | 130 | 50.00 |
| Computer Science; Engineering | 56 | 21.54 |
| Computer Science; Telecommunications | 22 | 8.46 |
| Engineering; Telecommunications | 12 | 4.62 |
| Computer Science; Engineering; Telecommunications | 8 | 3.08 |
| Engineering | 6 | 2.31 |
| Computer Science; Engineering; Information Science & Library Science; Government & Law; Telecommunications | 3 | 1.15 |
| Telecommunications | 3 | 1.15 |
| Automation & Control Systems; Computer Science | 2 | 0.77 |
| Computer Science; International Relations | 2 | 0.77 |
| Government & Law | 2 | 0.77 |
| Automation & Control Systems; Computer Science; Engineering | 1 | 0.38 |
| Automation & Control Systems; Engineering | 1 | 0.38 |
| Automation & Control Systems; Engineering; Materials Science | 1 | 0.38 |
| Computer Science; Criminology & Penology | 1 | 0.38 |
| Computer Science; Education & Educational Research | 1 | 0.38 |
| Computer Science; Engineering; Operations Research & Management Science | 1 | 0.38 |
| Computer Science; Medical Informatics | 1 | 0.38 |
| Computer Science; Operations Research & Management Science | 1 | 0.38 |
| Computer Science; Physics | 1 | 0.38 |
| Criminology & Penology | 1 | 0.38 |
| Engineering; Materials Science | 1 | 0.38 |
| Engineering; Medical Informatics | 1 | 0.38 |

Table 6 displays each research area in isolation and illustrates that the research Computer Science research contains majority of the published articles, featuring 88.46% of all publications. Engineering and Telecommunications area feature 35% and 18.46% of all publications respectively.

### 3.3 Institutions

This section discusses the number of publications attributed to different institutions to determine which institutions are prevalent within cloud forensics research.

Table 7 lists the institutions with at least three relevant publications. Combined they are credited for 44.83% of all publications. The University of Pretoria (South Africa), University of South Australia (Australia) and University College Dublin (Ireland) are the 3 leading institutions with 5.38%, 5% and 4.62% of total publications respectively. The list features 7 institutions from

Asia, 6 from Europe, 2 from North America and 1 each from Africa, Australia and The Middle east. The single country with the highest number of institutions is India with 3 institutions.

Table 7
List of Institutions

| Institution | Publications | Publications % | Country |
| --- | --- | --- | --- |
| University of Pretoria | 14 | 5.38 | South Africa |
| University of South Australia | 13 | 5.00 | Australia |
| University College Dublin | 12 | 4.62 | Ireland |
| Birla Institute of Technology and Science | 10 | 3.85 | India |
| Tsinghua University | 6 | 2.31 | Peoples R China |
| University of The Aegean | 6 | 2.31 | Greece |
| University of Alabama at Birmingham | 6 | 2.31 | USA |
| Military Technical Academy | 5 | 1.92 | Romania |
| University of New Orleans | 4 | 1.54 | USA |
| University of Plymouth | 4 | 1.54 | England |
| Central Police University | 3 | 1.15 | Taiwan |
| Cisco Systems Inc | 3 | 1.15 | India |
| G.H. Raisoni College of Engineering | 3 | 1.15 | India |
| Khalifa University | 3 | 1.15 | U Arab Emirates |
| Ministry of Public Security | 3 | 1.15 | Peoples R China |
| Nanjing University | 3 | 1.15 | Peoples R China |
| National University of Sciences and Technology | 3 | 1.15 | Pakistan |
| University of Derby | 3 | 1.15 | England |
| University of Naples Federico II | 3 | 1.15 | Italy |

3.4 Impact Journals

This section discusses the impact journals from the cloud forensics research. The findings from this section will help the researcher identify the best publication to promote their papers.

Table 8 lists the journal titles identified within this study and their citation and publication data for the period 2009 – 2016. Although Digital Investigation is the journal with the highest number of publications, it has a relatively low impact factor (1.211)
The Computer Surveys journal published by the Association for Computing Machinery (ACM) has the highest impact factor (5.243) and the highest average number of citations per paper (15.58). It also has the highest h-index which is a measure of predicting future scientific achievement proposed [48].Within the Computer Science, Theory & Methods citation reports category it is ranked 2 out of 105 whereas between the years 2002-2011 it was ranked at no. 1.
Table 9 lists the journals and books that have published at least three relevant articles.
The Digital Investigation journal is a clear leader with 10.39% of all publications. Digital Investigation is an international journal in digital forensics & incident response promoting innovations and advancement in the field.

**Table 8**
Impact journals

| Journal Title | P | C | CY | CI | H | IF | Q | % |
|---|---|---|---|---|---|---|---|---|
| ACM Computing Surveys | 357 | 5,561 | 695.12 | 15.58 | 33 | 5.243 | 1 | 0.86 |
| Future Generation Computer Systems the International Journal of Grid Computing and EScience | 611 | 6,431 | 803.88 | 10.53 | 28 | 2.430 | 1 | 0.86 |
| IEEE Transactions on Dependable and Secure Computing | 387 | 2,459 | 307.38 | 6.35 | 21 | 1.592 | 1 | 0.86 |
| Digital Investigation | 322 | 1,319 | 164.88 | 4.1 | 17 | 1.211 | 2 | 10.78 |
| Computer | 2,113 | 7,213 | 801.44 | 3.41 | 33 | 1.115 | 2 | 1.29 |
| Tsinghua Science and Technology | 228 | 379 | 75.8 | 1.66 | 8 | 1.063 | 4 | 1.72 |
| Journal of Internet Technology | 835 | 1,249 | 156.12 | 1.5 | 10 | 0.533 | 4 | 0.86 |
| Computer Law Security Review | 492 | 478 | 95.6 | 0.97 | 10 | 0.373 | 3 | 0.86 |

P, No. Publications; C, No. citations; CY, Average citations per year; CI, Average citations per item; H, h-index;
IF, Impact factor; Q, Quartile in category

**Table 9**
Top source titles

| Source Titles | No. Publications | % |
|---|---|---|
| Digital Investigation | 27 | 10.39 |
| Lecture Notes in Computer Science | 11 | 4.23 |
| IFIP Advances in Information and Communication Technology | 11 | 4.23 |
| Proceedings of The International Conference on Cloud Security Management | 7 | 2.69 |
| Advances in Digital Forensics XI | 4 | 1.54 |
| Communications in Computer and Information Science | 4 | 1.54 |
| Computers Security | 4 | 1.54 |
| Lecture Notes of the Institute for Computer Sciences Social Informatics and Telecommunications Engineering | 4 | 1.54 |
| Proceedings 10th International Conference on Availability Reliability and Security Ares 2015 | 4 | 1.54 |
| Tsinghua Science and Technology | 4 | 1.54 |
| 2013 Eighth International Workshop on Systematic Approaches to Digital Forensic Engineering SADFE | 3 | 1.15 |
| Advances in Digital Forensics VIII | 3 | 1.15 |
| Computer | 3 | 1.15 |
| Digital Forensics and Cyber Crime ICDF2C 2012 | 3 | 1.15 |
| IEEE Cloud Computing | 3 | 1.15 |
| IEEE International Advance Computing Conference | 3 | 1.15 |
| Information Security for South Africa | 3 | 1.15 |
| International Workshop on Systematic Approaches to Digital Forensic Engineering SADFE | 3 | 1.15 |
| Procedia Computer Science | 3 | 1.15 |
| Proceedings of 2016 11th International Conference on Availability Reliability and Security Ares 2016 | 3 | 1.15 |
| Proceedings of the 3rd International Conference on Cloud Security and Management ICCSM 2015 | 3 | 1.15 |
| Proceedings of The International Conference on Information Warfare and Security | 3 | 1.15 |

## 3.5 Highly-Cited Articles

This section discusses the most highly-cited articles within both journal and book publications. Figure 6 shows that 78.71% of all publications in this study do not yet have any citations. Only 5.94% of the 260 publications have greater than 10 citations.

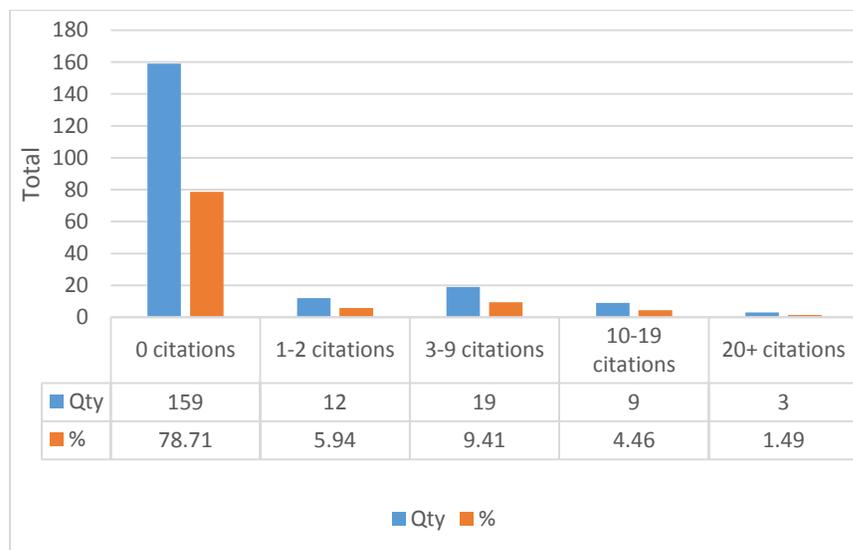

**Figure 6.** No. citations

Table 10 lists the articles that have received more than 10 citations between 2009 and 2016. It also includes details of the publication (journal or book series), published year and research area. The Computer Science research area comprises 89% of this list and the journal Digital Investigation is responsible for 57.89% all published articles. The top 3 cited articles were all published in 2012 which reinforces the idea that the earlier a work is published, the more it will be cited. The top 3 articles are high quality forensics procedural-based research papers [38] that are acknowledged by future authors due to their originality and value [49].

**Table 10**
Highly-cited articles

| Title | Times Cited | Publication | Year | Research Areas |
|---|---|---|---|---|
| Acquiring Forensic Evidence from Infrastructure-As-A-Service Cloud Computing: Exploring and Evaluating Tools, Trust, And Techniques | 44 | Digital Investigation | 2012 | Computer Science |
| An Integrated Conceptual Digital Forensic Framework for Cloud Computing | 43 | Digital Investigation | 2012 | Computer Science |
| Digital Forensic Investigation of Cloud Storage Services | 37 | Digital Investigation | 2012 | Computer Science |
| Cloud Forensics Definitions and Critical Criteria for Cloud Forensic Capability: An Overview of Survey Results | 31 | Digital Investigation | 2013 | Computer Science |
| Dropbox Analysis: Data Remnants on User Machines | 28 | Digital Investigation | 2013 | Computer Science |
| Cloud Storage Forensics: OwnCloud as A Case Study | 26 | Digital Investigation | 2013 | Computer Science |
| Efficient Audit Service Outsourcing for Data Integrity in Clouds | 25 | Journal of Systems and Software | 2012 | Computer Science |
| Design and Implementation of Frost: Digital Forensic Tools for The Openstack Cloud Computing Platform | 25 | Digital Investigation | 2013 | Computer Science |
| Digital Droplets: Microsoft Skydrive Forensic Data Remnants | 25 | Future Generation Computer Systems - The International Journal of Grid Computing and EScience | 2013 | Computer Science |
| Forensic Collection of Cloud Storage Data: Does the Act of Collection Result in Changes to The Data or Its Metadata? | 23 | Digital Investigation | 2013 | Computer Science |
| Google Drive: Forensic Analysis of Data Remnants | 21 | Journal of Network and Computer Applications | 2014 | Computer Science |
| Cloud Computing-Based Forensic Analysis for Collaborative Network Security Management System | 20 | Tsinghua Science and Technology | 2013 | Computer Science; Engineering |
| Cloud Computing and Its Implications for Cybercrime Investigations in Australia | 18 | Computer Law & Security Review | 2013 | Government & Law |
| Impacts of Increasing Volume of Digital Forensic Data: A Survey and Future Research Challenges | 15 | Digital Investigation | 2014 | Computer Science |
| A Survey of Information Security Incident Handling in The Cloud | 14 | Computers & Security | 2015 | Computer Science |
| Amazon Cloud Drive Forensic Analysis | 12 | Digital Investigation | 2013 | Computer Science |
| Distributed Filesystem Forensics: XtreemFS as A Case Study | 11 | Digital Investigation | 2014 | Computer Science |
| A Forensically Sound Adversary Model for Mobile Devices | 11 | Plos One | 2015 | Science & Technology - Other Topics |
| Overcast: Forensic Discovery in Cloud Environments | 10 | IMF 2009: 5th International Conference on IT Security Incident Management and IT Forensic | 2009 | Computer Science |

### 3.6 Keywords Frequency

This section discusses the use of author keywords and how it enables researchers to identify specific research [50]. The publications within this study contained 1172 keywords across 260 publications; 35 did not contain any keywords. The largest number of keywords for publications is 17; the average number of keywords is 4.52. The top author keywords and their relationship to their occurrence within the title are provided in Table 11.

**Table 11**
Relationship between title and author keywords

| Titles | Frequency | Keywords | Frequency |
| --- | --- | --- | --- |
| Cloud | 187 | Cloud computing | 104 |
| Forensic | 178 | Digital forensics | 78 |
| Forensics | 81 | Cloud forensics | 66 |
| Digital | 68 | Computer forensics | 20 |
| Digital forensic | 58 | Security | 19 |
| Cloud computing | 48 | Digital evidence | 12 |
| Analysis | 42 | Cloud storage | 11 |
| Cloud forensic | 34 | Cloud | 10 |
| Data | 34 | Forensics | 10 |
| Digital forensics | 33 | Digital investigation | 10 |
| Cloud forensics | 26 | Big data | 7 |
| Model | 23 | Network forensics | 7 |
| Evidence | 21 | Evidence | 6 |
| Security | 19 | Privacy | 6 |
| Challenges | 15 | Cybercrime | 6 |
| Forensic analysis | 13 | Cloud forensics challenges | 6 |
| Log | 13 | Virtualization | 6 |
| Forensic investigation | 12 | Digital | 5 |

The top 3 author keywords are "cloud computing", "digital forensics" and "cloud forensics". Within this study the top 3 author keywords have been included together in 19 publications; at least 2 of the top 3 keywords have featured in 46 publications. The mostly used keywords paper titles are "cloud", "forensic" and "forensics" with 187, 178 and 81 occurrences respectively. The author keyword "forensic" is included in just 4 publications but features highly in titles. Two examples of this are the publications "Cloud Manufacturing: Security, Privacy and Forensic Concerns" and "An Integrated Conceptual Digital Forensic Framework for Cloud Computing".

## 4 Challenges and Future Trends

This section discusses the limitations that a forensic investigator may face during examination within a cloud environment. Moreover, issues relating to data sovereignty, data confidentiality, and inadequacy of existing legislative and regulatory frameworks are elaborated [51].

## 4.1 Evidence Identification

Identification is the reporting of any malicious activity in the cloud such as illegal file storage or file deletion. The identification phase is initiated as a result of a complaint made by an individual or by a CSP authority that reports any misuse of the cloud [52]. The distributed nature of cloud computing makes evidence identification a difficult task. The first evidence collection issue that an investigator will encounter is of the system status and log files. Whereas this is not possible to collect in either a SaaS or PaaS model, it is possible in an IaaS cloud model where access is provided through a Virtual Machine (VM); within this model the VM behaves almost the same as an Actual Machine [53].

Data loss from volatile storage is the next issue facing a forensic investigator as all the client's data is volatile due to the high dependence on cloud computing. Also, due to the nature of cloud computing storage policies any evidence or stored data in the volatile storage will be removed, or deleted, if the criminal restarts or forces a power down of the computer.

Client-side evidence identification is another necessary step in computer forensic investigations that is usually not possible, especially in SaaS and PaaS models. In both models there are always some vital parts of evidence data that can be found on the client side interface (e.g. web browser temp data) [54]. Thus, the fragile and volatile nature of cloud environments require more attention and specialist techniques to ensure that the evidence data can be properly evaluated and isolated.

## 4.2 Legal Issues in the Cloud

Special care should be taken in from the outset to ensure that privacy of users are not violated during investigation of any criminal case [55]. Crimes involving cloud computing typically involve an accumulation or retention of data on a digital device (such as a mobile phone) that must be identified, preserved, analysed, and presented in a court of law [56]. Cloud data distribution within numerous data centres around the world creates jurisdictional issues relating to locating and seizing elusive evidential data [57]. Because of the nature of cloud computing, investigations require a co-operation between government agencies and law enforcement investigations from different countries, in addition to a collaboration of cloud service providers.

## 4.3 Data Collection and Preservation

Data collection in a computer forensic investigation is a significantly vital task and requires a physical acquisition for any forensic investigation. For example, within digital forensics the process of taking custody of any storage device (including hard disk) and then taking a bit-by-bit image for this device is one of the procedures that must be performed. This becomes a key issue in cloud computing as this step of the process is not possible due to the shared nature of the cloud environment. The investigator may have to contact the CSP for physical acquisition of data because these resources are distributed between numerous data centers, as previously discussed. Moreover, resources can be shared simultaneously among multiple cloud clients and can be constantly in use. The privacy of other client's data is therefore another issue faced in the seizure of physical evidence [58].

The data collection phase of cloud forensics should also consider the storage capacity for collecting evidence [59]. The amount of extracted data and the collected evidence would be greater than non-cloud digital forensics because of the wider nature of the cloud. The preservation of the evidence in a forensic investigation is vital to prove that an offence has been committed and how it relates to evidence can make it inadmissible. Another issue in evidence collection and preservation is the chain of custody which is the chronological documentation that shows how evidence was collected, preserved and analyzed [59]. Again, due to the cloud nature this attribute can violate digital forensic rules. To solve this challenge having a multifactor authentication method can prevent the perpetrator from claiming stolen authentication credentials [54].

### 4.4 Analysis and Presentation

Data analysis is another phase involving the analysis of collected data from different resource layers. In cloud computing this step has significant challenges because of the utilization of the intensive computation and massive data within cloud computing. This becomes an additional issue for cloud forensics investigation mainly due to the limitations in processing and examining vast amounts of data [60]. During forensics presentation, the judge or jury members may not fully understand the validity of evidence collected from the cloud, or comprehend what they are being told, or shown.

### 4.5 Future Trends

When considering the challenges that cloud computing environments offer there are several areas of future research that could be undertaken, for example: the evidence collection process and data volatility in SaaS and PaaS cloud models; chain of custody and privacy considerations for the seizure of physical evidence; jurisdiction and multi-agency/provider collaboration within cloud environments.

**Conclusion**

Cloud computing and the internet are interrelated and that makes them increasingly vulnerable to security threats. Digital forensic practitioners must extend their expertise and toolsets to conduct cloud examination. Moreover, cloud-based entities, CSPs and cloud customers must consider including built-in forensic capabilities to their platforms. In this paper the bibliometric methods were used to analyse cloud forensic research trends from 2009 until 2016. We presented criteria including publication statistics, citation distributions and statistics, regional and institutional productivity, research areas, impact journals, and keywords frequency. These criteria helped to uncover the global trends and significant areas in cloud forensic research. It is noticeable that the number of publications relating to cloud forensics has increased with an average annual growth rate of 218%. Asia had the largest number of publication in academic research followed by Europe and North America respectively. This paper findings provide researchers with better understanding of emerging trends in cloud forensic and help them to identify key areas for future research in this field.

**Response to reviewers**

Please find the response to each review item below.

Regards,

James Baldwin

**Reviewer 1:**

*1) Some charts did not provide with enough information like figure no. 1*

This table is provided to illustrate the growth in cloud adoption across a variety of sectors.

For clarity I have reworded the chart introduction text with the changes highlighted:- "According to the 2016 State of the Cloud Survey in which 1060 technical professionals representing a broad cross section of organizations were questioned [8], there has been an increase (from 2015) in the number of organisations utilizing the services of cloud providers. This change is illustrated in Figure 1."

The position of the chart has also been moved adjacent to the referring text.

*2) the paper text does not formatted well it should be justify.*

I have justified the entire paper.
Please note that this change is not highlighted on the actual paper itself.

*3) the whole paper does not formatted as journal requirements. please see the authors instruction.*

I have made several changes to the formatting:

- The chapter title page has been amended to include the author affiliation and email details at the foot of the page in a reduced font size.
- The extra spacing between the keywords and introduction has been removed
- The heading numbers have been amended e.g. from 1.   Introduction to **1   Introduction**
- Each section opening paragraph tab spacing has been removed and subsequent paragraphs have had the additional line space removed.

**Reviewer 2:**

*Please elaborate more on the lessons learned from this study, research challenges and future research directions.*

Added a new section 4.5. Future Trends and moved examples from the conclusion into this section, also adding additional examples.

*The organization and writing of the paper should be improved.*
*- Please provide a more precise explanation and example of SaaS, PaaS, and IaaS. In page 2, AWS is considered as both PaaS and IaaS, while I would consider it as IaaS.*

This section has been reworded slightly to provide more clarity with the changes highlighted : In the SaaS model an Application Service Provider (ASP) provides various applications over the Internet which eliminates the need for software and IT infrastructure (servers/databases etc) maintenance for the ASP customers [4]. The applications are accessed using a client browser interface. Google Apps, Yahoo Mail and CRM applications are all instances of SaaS. In the PaaS model, cloud infrastructure is owned and maintained by the provider; the customer is then able to deploy and configure applications into a provider managed framework and infrastructure [5]. Examples of PaaS are Google App Engine, Apprenda and Heroku. In the IaaS model resources are provided to the customer as virtualised resources e.g. Virtual Machines (VMs). Whereas the customer has full control over the operating system, the provider maintains control over the physical hardware. This allows for services to be scaled and billed in line with customer resource requirements [6]. Amazon Web Services (AWS), Microsoft Azure and Google Compute Engine (GCE) are examples of IaaS models.

*- Table 1 should be improved, some of the functionalities are not well explained, e.g., FTK.*

The functionalities have been rewritten to provide better explanation and new references 15-24 have been added for each tool.  The changes are highlighted below:

| Utilized tools | General/cloud-based tools | Functionality |
|---|---|---|
| FTK remote agent [15] | General | Remote drive and memory image acquisition; remote mounting |
| Encase remote agent [16] | General | Remote drive image acquisition |
| Snort [17] | General | Network traffic monitoring and packet logging |
| FROST [18] | Cloud-based | Digital forensics tools for the OpenStack cloud platform |
| OWADE [19] | Cloud-based | Reconstruction of browsing history and online credentials |
| CloudTrail [20] | Cloud-based | Logging in the AWS cloud |
| Wireshark [21] | General | Network traffic capture and analysis |
| Sleuthkit [22] | General | Forensic image analysis and data recovery |
| FTK imager [15] | General | Acquisition of memory and disk images |
| X-Ways [23] | General | Acquisition of live systems (Windows and Linux) |
| Encase e-discovery suite [24] | General | Drive image acquisition and offline examination |

*- What is the \* character here (page 4): 'cloud forensic\*', forensic\* and 'cloud investigation'?*

This section has been reworded and the changes are highlighted:
"We used various general search terms including 'cloud forensic\*', forensic\* (where \* denotes a wildcard character in the search term) and 'cloud investigation' in addition to specific search terms to include platforms, service models and deployment methods."

*Due to formatting and printing requirements please change Figure 2 to horizontal and elaborate more on the meaning of the arrows between the boxes.*

Figure 2 has been reformatted and I think it now shows a clearer process flow in a horizontal direction. It has also been inserted as an image.
The revised figure 2:

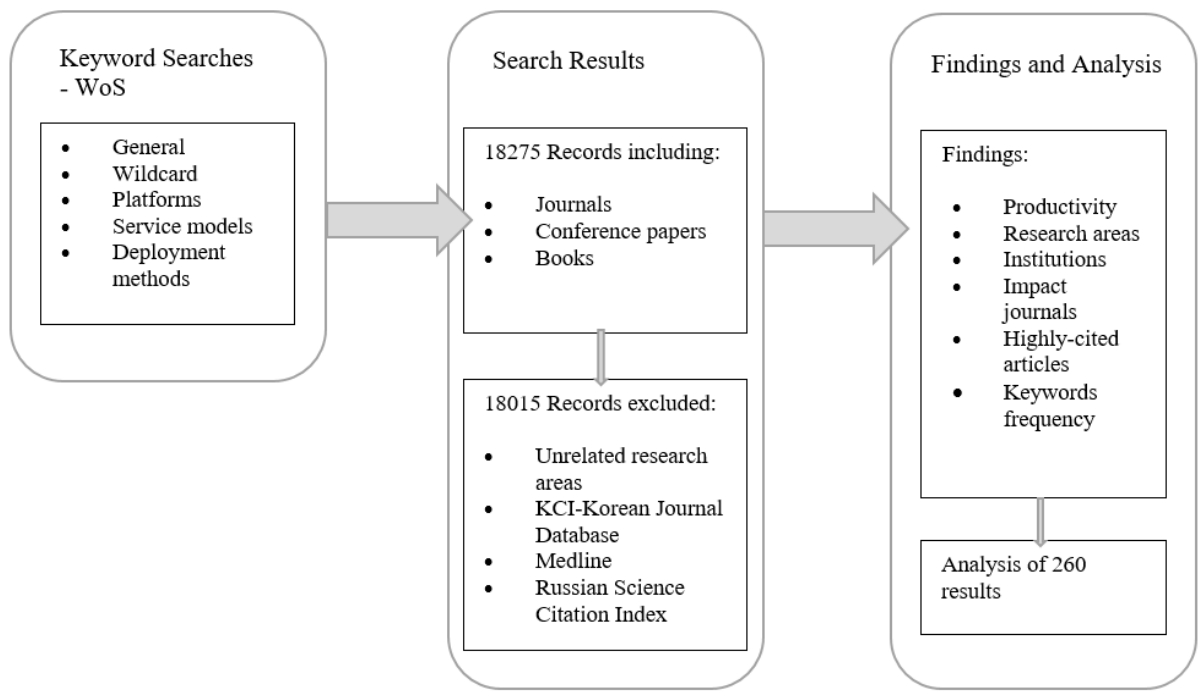

**Figure 2.** The data collection process